\documentclass[11pt,aps,onecolumn,pra,showpacs,amsmath,footinbib,superscriptaddress]{revtex4-1}
\usepackage{graphicx}
\usepackage{epstopdf}
\usepackage{graphics}
\usepackage{mathrsfs}
\usepackage{float}
\usepackage{psfrag}
\usepackage{amsmath,amssymb}
\usepackage{colordvi}
\usepackage{color}
\usepackage{verbatim}
\usepackage{soul}
\renewcommand{\a}{\alpha}
\renewcommand{\b}{\beta}

\renewcommand{\O}{\Omega}

\renewcommand{\l}{\lambda}

\renewcommand{\dag}{\dagger}

\newcommand{\bra}[1]{\left\langle{#1}\right|}
\newcommand{\ket}[1]{\left|{#1}\right\rangle}
\newcommand{\lr}[1]{\left\langle#1\right\rangle}

\newcommand{\be}{\begin{equation}}
\newcommand{\ee}{\end{equation}}
\newcommand{\ba}{\begin{eqnarray}}
\newcommand{\ea}{\end{eqnarray}}

\begin{document}
\title{Evidence for correlated states in a cluster of bosons with Rashba spin-orbit coupling}
\author{Zhihao Xu}
\affiliation{Institute of Theoretical Physics, Shanxi University, Taiyuan 030006, China}
\author{Zhenhua Yu}
\affiliation{Institute for Advanced Study, Tsinghua University,
Beijing 100084, China}
\email{huazhenyu2000@gmail.com}
\author{Shizhong Zhang}
\affiliation{Department of Physics and Center of Theoretical and
Computational Physics, The University of Hong Kong, Hong Kong,
China}
\email{shizhong@hku.hk}

\date{\today}
\begin{abstract}
We study the ground state properties of spin-half bosons subjected to the Rashba spin-orbit coupling in two dimensions. Due to the enhancement of the low energy density of states, it is expected that the effect of interaction becomes more important. After reviewing several possible ideal condensed states, we carry out an exact diagonalization calculation for a cluster of the bosons in the presence of strong spin-orbit coupling on a two-dimensional disk and reveal strong correlations in its ground state. 
We derive a low-energy effective Hamiltonian to understand how states with strong correlations become energetically more favorable than the ideal condensed states.
\end{abstract}

\maketitle

\section{Introduction}
A paradigm of traditional bosonic quantum liquid is that of Helium-4. At low temperature, Helium-4 is an inert element with no relevant internal degrees of freedom. At zero temperature, about $10$\% of the Helium atoms condense into the lowest energy state with zero momentum \cite{Svensson}. The effects of the inter-atomic interaction do not destroy the basic phenomena of Bose condensation which was first predicted for noninteracting bosons, even though significant depletion is resulted from inter-particle scattering \cite{Yang1962,Leggett}. Another general feature of Helium-4 in bulk is that it obeys Galilean invariance, which plays a crucial role in Landau's formulation of the two fluid model for liquid Helium-4~\cite{Landau,Khalatnikov}.

In 1995, Bose-Einstein condensation was realized in cold atomic gases. The observed atomic condensates are very close to an ``ideal" condensate for noninteracting bosons since the inter-atomic interactions are usually very weak. With the advent of synthetic spin-orbit coupling in cold atomic gases, the study of bosonic quantum liquid is greatly expanded in several aspects~\cite{Dalibard,Goldman,Galitski,Congjun,HuiZhai1,Zhang}. First of all, alkali elements like $^{87}$Rb usually have multiple internal hyperfine states, and in an optical trap all of them can be active and this leads to multitude of quantum phases in the so-called spinor condensate~\cite{Ueda,Ueda1,Ho1998,ohmi1998,Stenger}. Secondly, the inclusion of spin-orbit coupling further introduces the coupling between the spin and momentum degrees of freedom and significantly modifies the single particle dispersion relation. In the experimentally realized case of spin-orbit coupling along one-dimension~\cite{Spielman1,Spielman2,Spielman3,JYZhang,JWPan,Zwierlein,JZhang1,JZhang2,JZhang3,Hamner,Williams,Beeler,LeBlanc,Olson,Khamehchi}, this coupling leads to the so-called stripe and plane wave condensate in homogeneous systems and in harmonic traps~\cite{Shizhong, HuiZhai,Ozawa,YPZhang, YunLi1, Santos2011, Santos2014} and the associated tricritical point~\cite{YunLi2}. Thirdly, for certain symmetric spin-orbit coupling, of Rashba or Weyl form, the low energy density of states is significantly enhanced such that the effects of interaction have drastic effects on the existence of Bose condensate and could in fact destroy its existence~\cite{XiaolingCui,QiZhou}. Lastly, the inclusion of spin-orbit coupling breaks the Galilean invariance which makes the construction of two-fluid model much more involved~\cite{QZhu,WeiZheng,Valiente,HPu}.

In this paper, we consider spin-half bosons subjected to the Rashba spin-orbit coupling in two-dimensions. We will concentrate on the interplay between the enhanced single particle ground state degeneracy and the effects of strong repulsive interactions. In an infinite system, the Rashba spin-orbit coupling gives rise to a ring of infinitely degenerate single particle states. We show that in the truncated Hilbert space spanned by the single particle states on the ring, trial wave-functions for two and four bosons with strong correlations built-in can have lower interaction energy than various ideal condensed states. However, for strong inter-particle interaction, it is also important to take into account the transverse excitations away from the degenerate ring. We consider explicitly a disk of radius $R$ and carry out exact diagonalization calculations for a cluster of bosons in this finite disk. We find numerical evidences for the correlated nature of the ground state. We derive a low-energy effective Hamiltonian in the strong spin-orbit coupling limit and use it to understand why the ground state of the interacting bosons exhibit strong correlations, analogous to one-dimensional Mott insulators and quantum Hall states~\cite{Lee2004,Lee2005}.

The paper is organized as follows. In Sec.~\ref{general_setup}, we discuss the general feature of a spin-half boson subjected to Rashba spin-orbit coupling in two-dimensional space and obtain its single particle eigen-spectrum and density of states for later discussions. In Sec.~\ref{mft_condensate}, we list a host of possible condensate states and calculate their energies.  In Sec.~\ref{two-body_solution}, we discuss a trial two-body wave-functions that totally avoid interactions. In Sec.~\ref{cs4particle}, we discuss the possibilities of correlated ground states in the four-particle case and show that the proposed correlated state is lower in energy than the best condensate states considered in Sec.~\ref{mft_condensate}. All the above discussions are based on trial wave functions constructed from the lowest degenerate ring. We then highlight the inadequacy of constructing trial states only from the degenerate ground states on the ring without considering transverse excitation (finite kinetic energy) when interaction becomes strong. In Sec.\ref{csnumerical}, we take into account the transverse excitation and perform exact diagonalization calculation of a cluster of bosons on a disk. To understand the result obtained, we derive an effective many-body Hamiltonian in the limit of strong spin-orbit coupling in Sec.\ref{cseffHam}. Two appendices (Sec.\ref{app}) are given. In Sec.\ref{app1}, we discuss the exact solution of single particle states on a disk and in Sec.\ref{app2}, we discuss the detailed construction of the low energy effective Hamiltonian.

\section{General Setup}
\label{general_setup}
\begin{figure}
\begin{center}
\includegraphics[width=1 \textwidth]{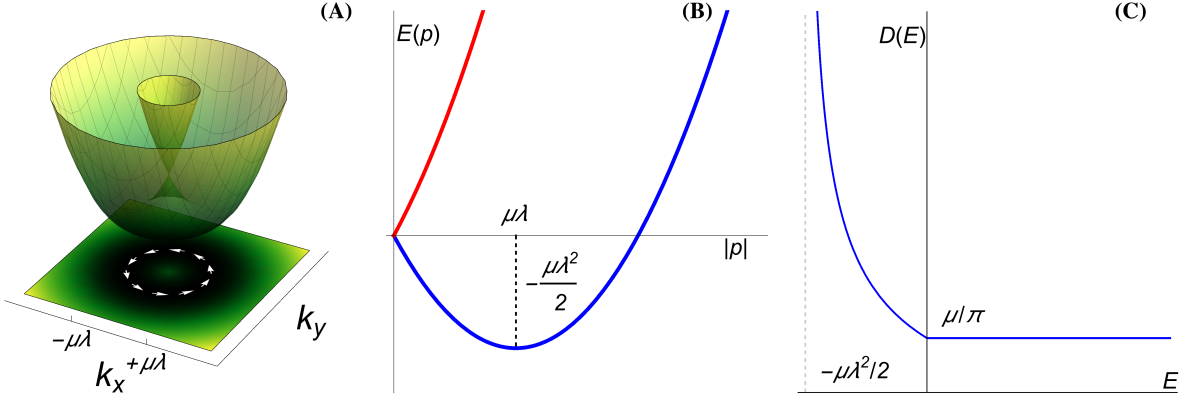}
\end{center}
\caption{(Color online) (A) Three-dimensional plot of single particle spectrum with Rashba spin-orbit coupling in two dimensions. The locus of lowest energy states lies in the circle with $|{\bf p}|=\mu\l$ and its spin direction is indicated in (A). The two branches of the single particle spectrum touch at $k_x=k_y=0$ at $E=0$. (B) A cut of the energy spectrum along one angular direction. It is shown that the lowest energy is given by $E_{\rm min}=-\mu\l^2/2$ at $|{\bf p}|=\mu\l$. The curvature of the lower branch at the minimal energy is given by $1/\mu$ and the dispersion becomes sharper as $\mu\to 0$. (C) The density of states $D(E)$ resembles one-dimensional system with characteristic $\sim 1/\sqrt{E+\mu\l^2/2}$ divergence for negative energy, and remains a constant for $E>0$, the ususal density of states in two-dimensions.}
\label{Figband}
\end{figure}

In the presence of the Rashba spin-orbit coupling, the single-particle Hamiltonian in two dimensions (2D) is given by
\begin{equation}
\label{eq1}
H_{\mathrm{kin}}=\frac{\hat{p}_{x}^{2}+\hat{p}_{y}^{2}}{2\mu}+\lambda(\sigma_{x}\hat{p}_y-\sigma_{y}\hat{p}_x),
\end{equation}
where $\hat{p}_{x(y)}$ is the momentum along $x$($y$)-direction, $\mu$ is
the mass of the boson, $\lambda$ is the strength of the spin-orbit coupling
and $\sigma_{x(y)}$ is the $x$($y$)-Pauli matrix acting on two internal states of a boson.
There are two branches of single-particle eigenstates
\begin{equation}
\label{eq2}
\chi_{\pm,\mathbf p}(\mathbf r)=\frac{1}{\sqrt{2\O}}\left(
\begin{array}{c}
1 \\
\pm i z_{\mathbf p}
\end{array}%
\right)e^{i \mathbf{p} \cdot \mathbf{r}},
\end{equation}
whose eigenvalues are $\epsilon_{\pm,p}=p^{2}/{2\mu}\pm\lambda|\mathbf{p}|$ [see Fig.~\ref{Figband} (A-B)].
Here $\O$ is the area of the two-dimensional system, $z_{\bf p}=e^{i\varphi_{\bf p}}$ with $\varphi_{\bf p}$ being
the polar angle of $\mathbf{p}$, i.e.,
$p_{x}=p\cos{\varphi_{\mathbf{p}}}$ and $p_{y}=p\sin{\varphi_{\mathbf{p}}}$. The
minimum of the lower Rashba band occurs at $|\mathbf{p}|=\mu\lambda$.  Note that
$L_{z}+S_{z}$ is a good quantum number with $\mathbf S=\boldsymbol\sigma/2$. Using the fact that $\exp(i{\bf p}\cdot{\bf r})=\sum_m i^m J_m(pr)\exp(im\theta)$, where $\theta$ is the angle between ${\bf p}$ and ${\bf r}$, it is straightforward to show that the alternative spinor wave-functions (not normalized)
 \begin{equation}
\label{eq3}
\chi_{\pm,p,m}= \left(
\begin{array}{c}
e^{i m\varphi_{\bf r}} J_{m}(pr)\\
\pm e^{i (m+1) \varphi_{\bf r}} J_{m+1}(pr)
\end{array}%
\right)
\end{equation}
satisfy
 \begin{eqnarray}
\label{eq4}
H\chi_{\pm,p,m}&=&\left(\frac{p^2}{2\mu}\pm \lambda|\mathbf{p}|\right)\chi_{\pm,p,m}, \\
(L_z+S_z)\chi_{\pm,p,m}&=&(m+1/2)\chi_{\pm,p,m},
\end{eqnarray}
where $J_m(x)$ is the $m$-th order Bessel function, and $\varphi_{r}$ is the polar angle of $\mathbf{r}$ with respect to ${\bf p}$.

%-----------------------figure--------------------------------------------------
%-----------------------figure--------------------------------------------------

One consequence of the inclusion of the Rashba spin-orbit interaction is that the low energy density of states is significantly enhanced; see Fig.~\ref{Figband} (C). The density of states $D(E)$ can be conveniently computed by counting the number of states with energy below $E$, namely $N(E)=\sum_{\bf k}[\theta(E-\epsilon_{-,k})+\theta(E-\epsilon_{+,k})]$, where $\theta(x)$ is the Heaveside step function, and $D(E)=dN(E)/\O dE$ gives 
\begin{align}
D(E)=
   \begin{cases}
     \frac{\mu^{3/2}\lambda}{\sqrt{2}\pi} \frac{1}{\sqrt{E+\mu\l^2/2}}, &\text{if $E < 0$;}\\
     \frac{\mu}{\pi}, &\text{if $E > 0$}.
   \end{cases}
\end{align}
Namely, due to the spin-orbit coupling, the low energy ($E\to-\mu\l^2/2$) density of states diverges as in the one-dimensional case, while those for $E>0$ retains the usual two-dimensional constant density of states.

In the following, we assume that the repulsive interactions between the bosons can be modeled by the contact psuedopotential
\begin{equation}
\label{eq5}
V(\mathbf{r}_1-\mathbf{r}_2)=(U+U_s\boldsymbol\sigma_1 \cdot \boldsymbol\sigma_2)\delta(\mathbf{r}_1-\mathbf{r}_2),
\end{equation}
where $\boldsymbol\sigma_{1,2}$ are the Pauli matrices of the two particles, and $U$ and $U_s$ are
positive interaction coupling constants. $U$ describes the spin-independent density-density interactions, while $U_s$ describes the spin-dependent spin exchange interaction. Due to Bose statistics, the contact psuedopotential is nonzero only when two bosons are in a spin triplet state; $V(\mathbf r)$ in Eq.~(\ref{eq5}) is equivalent to $\widetilde U\delta(\mathbf r)$ with $\widetilde U=U+U_s$. We should emphasise here that Eq.~(\ref{eq5}) is not the most general interaction between atoms and in fact, only approximate the real form of the interaction between $^{87}$Rb. However, we expect that such a simplified choice of interaction would not affect the nature of the correlated ground states which we discuss later, as their nature is rooted in the degeneracy brought about by the spin-orbit coupling.

\section{Mean field condensate states}
\label{mft_condensate}
Before we discuss the possibility of correlated ground states, let us first list a few candidate condensate states built from the lowest single particle states $|\mathbf p|=\mu\l$. For an ideal condensate of $N$ bosons, the trial wave-function can be written as
\begin{equation}
\label{eq17}
|\mathrm{Cond}\rangle=\frac{(\phi_0^{\dagger})^{N}}{\sqrt{N!}}|0\rangle,
\end{equation}
with $\phi_0^{\dagger}=\sum_{|\mathbf{p}|=\mu\lambda}c_{\mathbf{p}}
a_{\mathbf{p}}^{\dagger}$ and $a_{\mathbf{p}}^{\dagger}$ the lower Rashba branch creation operators. The expectation value of the interaction energy is given by ($z_{\mathbf{p}}=\exp(i\varphi_{\bf p})$)
\begin{eqnarray}
\label{eq18}
& &\langle\mathrm{Cond}|H_{\mathrm{int}}|\mathrm{Cond}\rangle \notag \\
&=&\frac{1}{2N!}\sum_{\mathbf{p}_1,\mathbf{p}_2,\mathbf{k}_1,\mathbf{k}_2}
\langle \mathbf{p}_1,\mathbf{p}_2|V|\mathbf{k}_2,\mathbf{k}_1\rangle
\langle 0|(\phi_0)^{N}a_{\mathbf{p}_1}^{\dagger}a_{\mathbf{p}_2}^{\dagger}a_{\mathbf{k}_2}a_{\mathbf{k}_1}(\phi_0^{\dagger})^{N}|0\rangle \notag \\
&=&\frac{N(N-1)}{2}\Big\{\frac{1}{4\O}\sum_{\mathbf{p}_1,\mathbf{k}_1}\left[(U+U_s)+2(U-3U_s)z_{\mathbf{p}_1}^{*}z_{\mathbf{k}_1}+(U+U_s)(z_{\mathbf{p}_1}^{*}z_{\mathbf{k}_1})^2\right]c_{\mathbf{p}_1}^{*}c_{-\mathbf{p}_1}^{*}c_{-\mathbf{k}_1}c_{\mathbf{k}_1} \notag \\
& &+\frac{U+U_s}{4\O}\sum_{\mathbf{p}_1,\mathbf{p}_2}[6+z_{\mathbf{p}_1}^{*}z_{\mathbf{p}_2}+z_{\mathbf{p}_1}z_{\mathbf{p}_2}^{*}]|c_{\mathbf{p}_1}|^2|\mathbf{p}_2|^2-\frac{U+U_s}{\O}\sum_{\mathbf{p}_1}|c_{\mathbf{p}_1}|^2|c_{-\mathbf{p}_1}|^2\Big\}.
\end{eqnarray}
We substitute different forms of $c_{\mathbf{p}}$, and the corresponding interaction energy expectation values, following the usual scaling $\sim \widetilde U N^2/\Omega$ expected for ideal condensates when $N$ is big, are listed in Table \ref{Tab1}. For the case of $c_{\mathbf{p}}=\frac{1}{\sqrt{2\pi}}[\sin{\alpha}e^{-in\varphi_\mathbf{p}}+\cos{\alpha}e^{i \phi}e^{-im\varphi_\mathbf{p}}]$ with $m \neq n$, it can be shown that the energy is lowest when $\alpha=0,\pi/2$
which is shown in Table \ref{Tab1}.

\begin{table}[!hbp]
\caption{Interaction energy expectation values of different condensate trial wave-functions.}
\begin{tabular}{|c|c|c|}
\hline
Case & $c_{\mathbf{p}}$ & $\langle\mathrm{Cond}|H_{\mathrm{int}}|\mathrm{Cond}\rangle$ \\
\hline
{I} & $\frac{1} {\sqrt{2\pi}}$ & $\frac{N(N-1)}{2}\frac{\widetilde U}{4\O}(7-\frac{2}{\pi})$  \\
\hline
{ II} & $\delta(\varphi_\mathbf{p})$ & $\frac{N(N-1)}{2}\frac{8\widetilde U}{4\O}$ \\
\hline
{III} &  $\frac{1}{\sqrt{2}}[\delta(\varphi_\mathbf{p})+e^{i\alpha}\delta(\varphi_\mathbf{p}-\pi)]$ &
$\frac{N(N-1)}{2}\frac{6\widetilde U}{4\O} $\\
\hline
{IV} & $\frac{1}{\sqrt{2\pi}}e^{in\varphi_\mathbf{p}},$ for $n\neq0,-1$ & $\frac{N(N-1)}{2}\frac{\widetilde U}{4\O}(6-\frac{2}{\pi})$ \\
\hline
{V} & $\frac{1}{\sqrt{2\pi}}e^{-i\varphi_\mathbf{p}}$ & $\frac{N(N-1)}{2}\frac{\widetilde U}{4\O}(7-\frac{2}{\pi})$ \\
\hline
\end{tabular}\label{Tab1}
\end{table}

\section{Few-body trial wave-functions}
To demonstrate how the bosons can take advantage of the degeneracy of the lower Rashba branch at $|\mathbf p|=\mu\lambda$ to lower their interaction energy than an ideal condensate, we construct explicit trial wave-functions with correlations built in for two and four bosons.

\subsection{Two-body}
\label{two-body_solution}
We write down the following two-body wave-function
\begin{equation}
\label{eq:2bodyexact}
\psi_{0}=\sum_{|\mathbf{p}|=\mu\lambda}A_{\mathbf{p}}\left\{\left[
\begin{array}{c}
1\\
iz_{\mathbf{p}}
\end{array}%
\right]_1\otimes\left[
\begin{array}{c}
1\\
iz_{-\mathbf{p}}
\end{array}%
\right]_2e^{-i\mathbf{p}\cdot(\mathbf{r}_1-\mathbf{r}_2)}+\left[
\begin{array}{c}
1\\
iz_{-\mathbf{p}}
\end{array}%
\right]_1\otimes\left[
\begin{array}{c}
1\\
iz_{\mathbf{p}}
\end{array}%
\right]_2e^{i\mathbf{p}\cdot(\mathbf{r}_1-\mathbf{r}_2)}
\right\},
\end{equation}
where $A_{\mathbf{p}}$ is the amplitude of the wave-function to be determined by minimizing the interaction energy. Since we have constrained $|\mathbf p|=\mu\lambda$, the kinetic energy is already at its minimum. It turns out that by choosing appropriate $A_{\mathbf{p}}\sim e^{i2n\varphi_\mathbf{p}}$, with $n\neq 0,1$, one can show that $\psi_0(\mathbf{r}_1=\mathbf{r}_2)=0$. Due to the contact nature of the interaction potential, this means that Eq.~(\ref{eq:2bodyexact}) is an {\it exact} eigenstate of two interacting bosons. 
The wave-function (\ref{eq:2bodyexact}) is the same as the one found in Ref.~\cite{QiZhou}. Previously Refs.~\cite{Ozawa} and~\cite{Sarang2011} have also shown that two-body scattering states in the lower Rashba branch can have zero interaction energy once the bare interactions are properly renormalised.

The fact that there exists a great many choice of $n$ such that exact two-body states without interaction can be constructed raises the interesting possibility of many-body correlated states, which,  while still consists of states with $|\mathbf p|=\mu\lambda$, avoids the interaction energy by correlating bosons in the manner as embodied in the two-body wave-function, Eq.~(\ref{eq:2bodyexact}). Unfortunately, this straightforward generalization from two-body to many-body could not be consistently carried out, since Eq.~(\ref{eq:2bodyexact}) correlates two bosons on the opposite side of the degenerate circle, and for a many-body system, it is impossible to achieve for any arbitrary pair. This difficulty can be most easily seen in the four bosons case to be discussed below.

\subsection{Four-body}
\label{cs4particle}

%To show that a correlated trial state can be energetically more favorable than \textbf{an ideal condensed state which usually occurs in the presence of weak interactions}, 
If we try to build explicitly the two-body correlation in the four boson case, an trial wave-function can be constructed
\begin{equation}
\label{eq19}
|\mathrm{Corr}\rangle=\sum_{|\mathbf p|,|\mathbf k|=\mu\l}A_{\mathbf p}a_{\mathbf p}^{\dagger}a_{-\mathbf p}^{\dagger}B_{\mathbf k}a_{\mathbf k}^{\dagger}a_{-\mathbf k}^{\dagger}|0\rangle,
\end{equation}
with $A_{\mathbf p}=Ae^{i2n\varphi_\mathbf p}$ and $B_{\mathbf k}=Be^{i2m\varphi_\mathbf k}$. It is clear that we are simply building into the four particle wave-function the correlations that we have identified in the two-body case. If $m=n$, Eq.~(\ref{eq19}) can be viewed as a condensate of boson pairs, which is the same as the one used in Ref.~\cite{Nozieres} to study the condensation of paired bosons.
The normalization is $\langle\mathrm{Corr}|
\mathrm{Corr}\rangle=[(1+\delta_{mn})\pi^2+2\pi]A^2B^2$.
\begin{comment}
To calculate the interaction energy, it is convenient to use the symmetrized interaction matrix elements
\begin{equation}
\label{eq20}
\langle \mathbf{p}_1,\mathbf{p}_2|V|\mathbf{k}_2,\mathbf{k}_1\rangle_{s}=\langle \mathbf{p}_1,\mathbf{p}_2|V|\mathbf{k}_2,\mathbf{k}_1\rangle+\langle \mathbf{p}_2,\mathbf{p}_1|V|\mathbf{k}_2,\mathbf{k}_1\rangle+\langle \mathbf{p}_1,\mathbf{p}_2|V|\mathbf{k}_1,\mathbf{k}_2\rangle+\langle \mathbf{p}_2,\mathbf{p}_1|V|\mathbf{k}_1,\mathbf{k}_2\rangle.
\end{equation}
When the states are constrained on the ring, we have
 \begin{equation}
\label{eq21}
\langle \mathbf{p}_1,\mathbf{p}_2|V|\mathbf{k}_2,\mathbf{k}_1\rangle_{s}=\frac{f+g}{\O}\delta_{\mathbf{p}_1+\mathbf{p}_2-\mathbf{k}_1-\mathbf{k}_2}(2+z_{\mathbf{p}_1}^{*}z_{\mathbf{k}_1}+z_{\mathbf{p}_2}^{*}z_{\mathbf{k}_2}+z_{\mathbf{p}_1}^{*}z_{\mathbf{k}_2}+z_{\mathbf{p}_2}^{*}z_{\mathbf{k}_1}+2z_{\mathbf{p}_1}^{*}z_{\mathbf{k}_1}z_{\mathbf{p}_2}^{*}z_{\mathbf{k}_2}).
\end{equation}
Therefore we have
\end{comment}
Direct evaluation yields
 \begin{eqnarray}
\label{eq22}
\langle\mathrm{Corr}|H_{\mathrm{int}}|\mathrm{Corr}\rangle
&=& \frac{12\widetilde U}{\O}(1+\delta_{mn})\pi^2A^2B^2.
\end{eqnarray}
Thus the expectation value of the interaction energy for the correlated wave-function is
 \begin{equation}
\label{eq23}
\frac{\langle\mathrm{Corr}|H_{\mathrm{int}}|\mathrm{Corr}\rangle}{\langle\mathrm{Corr}|
\mathrm{Corr}\rangle}=\frac{12\widetilde U}{\O}\frac{1+\delta_{mn}}{1+\delta_{mn}+2/\pi},
\end{equation}
which is approximately $7.33\times\widetilde U/\O$ for $m\neq n$ and $9.1\times\widetilde U/\O$ for $m=n$, compared with the
lowest interaction energy of an ideal condensate of Eq.~(\ref{eq17}) given in Tab.~\ref{Tab1}, Case IV with $N=4$, $(9-3/\pi)\widetilde U/\O\approx 8.04\times\widetilde U/\O$. Thus for four bosons, Eq.~(\ref{eq19}) with $m\neq n$ has lower interaction energy than the ideal trial condensed states. A super-fragmented trial state was proposed in Ref.~\cite{QiZhou}, whose interaction energy, if evaluated with our assumed interaction potential for $N=4$, would be $3.44\tilde U/\Omega$, even lower. This comparison indicates that different ways to correlate the bosons can suppress the interaction energy to different degrees.

Despite the effort to build in two-body correlation, the condensate states and the trial correlated wave functions for four bosons have at least two common features. Firstly, they are all built from the lowest energy states on the degenerate ring and secondly their energies are all proportional to $1/\Omega$ and coupling constant $\widetilde{U}$. When $\widetilde{U}=0$, all bosons will condense in the lowest single particle state. As one increases $\widetilde{U}$ from zero but still keeps it small compared to the single particle excitation energy, it is expected that the system falls in the mean field regime, as discussed above. However, as $\widetilde{U}$ increases further, the weakly dispersing lowest Rashba band becomes very important. One way to see this is that if one starts with a finite system with area $\Omega$, the single particle states in the lower Rashba branch with momentum $p-\mu\lambda\lesssim 1/\sqrt\Omega$ differ only by energy of order of $1/\Omega$ with the states in the degenerate ring. On the other hand, the apparent scale of the interaction energy is also $1/\Omega$ (cf.~Eq.~(\ref{eq23})). This indicates that, at least in the strong interaction regime, it is important to consider the transverse excitations, and their effects have to be taken into account in the construction of effective Hamiltonian (see Sec. \ref{cseffHam}). The above considerations prompt us to study the bosons in a disk of finite size $\Omega$ in the next section, where we are able to treat both the kinetic and interaction energies on the same footing.

\section{Small Clusters in a Disk.}
\label{csnumerical}

Correlated ground states for a homogeneous interacting Bose gas with the Rashba spin-orbit coupling~\cite{XiaolingCui, QiZhou} have been explored by means of fermionization~\cite{Glazman} and by vortex attachement~\cite{Scarola}. Numerical calculations have also revealed the correlated nature of the system in a harmonic trap~\cite{HuiHu2012}.

To understand how strong spin-orbit coupling can help the bosons to suppress interaction energy, we study a cluster of bosons with the Rashba spin-orbit coupling in a two-dimensional disk of radius $R$ with a hard wall boundary condition, while earlier works has focused on two-dimensional harmonic trap confinement~\cite{Ramachandhran1,Ramachandhran2,HuiHu1,HuiHu3,XiaolingCui}.
%To understand why correlated states can have lower energy than condensed states, we consider the limit $\lambda\to \infty$ and extract the leading term of the interaction Hamiltonian as an effective low energy Hamiltonian.  
%For the convenience of calculation, we confine the bosons 
In this case, the single particle eigen-wave-functions with eigenenergies $E_{nm}$ have the generic form
\begin{eqnarray}
\label{eq26}
\Phi_{nm}(r,\varphi)&=&\left(
\begin{array}{c}
f_{nm}(r) e^{im\varphi}\\
\xi_{nm+1}(r)e^{i(m+1)\varphi}
\end{array}%
\right),
\end{eqnarray}
where $m+1/2$ is the eigenvalue of $S_z+L_z$, and $n$ is the radial quantum number determined when $\Phi_{nm}(R,\varphi)=0$ is imposed on the wave-function in the radial direction. The detailed expressions of $f_{nm}(r)$ and $\xi_{nm+1}(r)$ in terms of Bessel functions are given in Appendix \ref{app1}. In the following, we use $x\equiv\sqrt{2\mu^2\l^2R^2}$ to quantify the spin-orbit coupling strength.

%-----------------------figure--------------------------------------------------
\begin{figure}[h]
\begin{center}
\includegraphics[width=1 \textwidth]{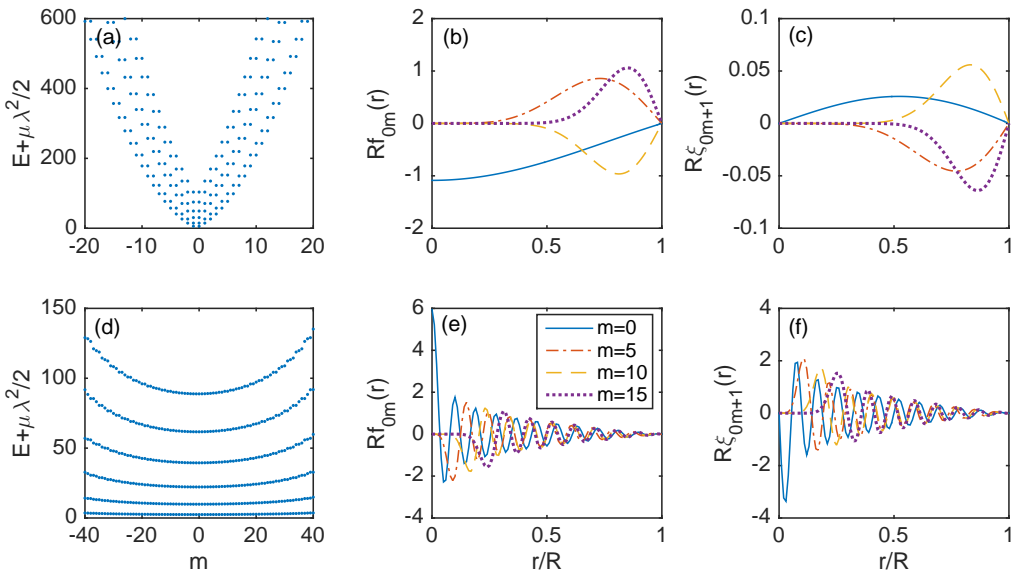}
\end{center}
\caption{(Color online) (a) Energy spectrum $E_{nm}$ for single-particle states at small spin-orbit
coupling strength ($x=0.1$) with radial quantum number $n=0\to5$ (we take $1/2\mu R^2$ as energy units); (b), (c) wave-function $f_{0m}(r)$
 and $\xi_{0m+1}(r)$ of single-particle states in the $n=0$ manifold at $x=0.1$ change with $r$; (d) Energy spectrum $E_{nm}$ for single-particle states at large spin-orbit
coupling strength ($x=100$) with radial quantum number $n=0\to5$; (e), (f) wave-function $f_{0m}(r)$
and $\xi_{0m+1}(r)$ of single-particle states in the $n=0$ manifold at $x=100$ change with $r$. In (b),(c), (e) and (f) solid
line is for $m=0$, dash-dotted line is for $m=5$,  dashed line is for $m=10$ and dotted line is for $m=15$.}
\label{Figspdisk}
\end{figure}
%-----------------------figure--------------------------------------------------

Figure \ref{Figspdisk} shows the energy spectrum $E_{nm}$ and the associated ground state wave-functions $f_{0m}(r)$ and $\xi_{0m+1}(r)$ ($n=0$) for both small (a-c) and large (d-f) spin-orbit coupling strengths $x$. We take $1/2\mu R^2$ as energy units throughout. 
In Figure \ref{Figspdisk} (a), the energy spectrum for small spin-orbit coupling ($x=0.1$) is strongly
dispersive. The small spin-orbit coupling can be regarded as a perturbation to the usual quadratic kinetic energy. However, when the spin-orbit coupling strength is large enough such as $x=100$ as shown in Figure \ref{Figspdisk} (d), the energy spectrum segregates into distinct energy bands and the lower energy bands corresponding to smaller $n$ are more flat. For $x=100$, the band bottom of the lowest energy band is found numerically to be at energy about $2.47\times 1/2\mu R^2-\mu\lambda^2/2$.
As an example, the energies
for $n=0$ are essentially non-dispersive for the states with $m\in[-40,40]$. In this limit, the quadratic kinetic
energy can be viewed as a perturbation to the spin-orbit coupling. Figure \ref{Figspdisk} (b) and (c) show the wave
functions $f_{0m}(r)$ and $\xi_{0m+1}(r)$ versus $r$ for the lowest energy band $n=0$. For a larger spin-orbit coupling strength ($x=100$), the wave-functions exhibit increased oscillations compared those with a small spin-orbit coupling strength. With increasing $|m|$, the weight of the wave-functions moves towards $r=R$. Here we would like to contrast the behavior of $f_{0m}(r)$ and $\xi_{0m+1}(r)$ with that in the quantum Hall problem, where the lowest Landau level wave-functions in the symmetric gauge are localized within distance $\ell$ at a set of concentric ring with radius $\sqrt{2m}\ell$, where $\ell$ is the magnetic length. In contrast, the wave-functions $f_{0m}(r)$ and $\xi_{0m+1}(r)$ are more spread out and with considerable oscillations for large spin-orbit coupling.

When the spin-orbit coupling strength $x$ is large [Fig.~\ref{Figspdisk} (d)], the $n=0$ energy
band is nearly flat and does not overlap with the higher bands for a large range of $m$ [cf.~Figure \ref{Figspdisk}(d)], we can project the Hilbert space into the lowest band $n=0$. As a result, we shall omit the radial quantum index $n$. The low energy physics can be described by the Hamiltonian $H=H_{\mathrm{kin}}+H_{\mathrm{int}}$ with $H_{\mathrm{kin}}=E_{m}a_m^{\dagger}a_m$, where $a_m^\dag$ is the creation operator for states with wave-function $\Phi_{0m}({\bf r})$. The interaction Hamiltonian is given by
\begin{align}
\label{eq27}
H_{\mathrm{int}}=&\sum_{m_1,m_2,m_3,m_4}V(m_1,m_2,m_3,m_4)a_{m_1}^{\dagger}a_{m_2}^{\dagger}a_{m_3}a_{m_4}\delta_{m_1+m_2-m_3-m_4},
\end{align}
where the matrix elements $V(m_1,m_2,m_3,m_4)$ are given by
\begin{align}
V(m_1,m_2,m_3,m_4)=\pi \widetilde U\int_{0}^{R}d r r[f^{*}_{m_1}(r)f_{m_4}(r)+\xi^{*}_{m_1+1}(r)\xi_{m_4+1}(r)][f^{*}_{m_2}(r)f_{m_3}(r)+\xi^{*}_{m_2+1}(r)\xi_{m_3+1}(r)].
\end{align}
It is worth to mention that the lowest band approximation we take here breaks down whenever the average energy of a boson (while the interaction energy may be small, the kinetic energy and thus the total energy can be still quite large) is comparable to the band gap such that higher bands are substantially populated.

\subsection{Exact Diagonalization}
We use the above Hamiltonian $H=H_\text{kin}+H_\text{int}$ projected to the lowest band to solve the states of $N$ bosons in the strong Rashba spin-orbit coupling limit by exact diagonalization. In the lowest band, we truncate the single particle states up to the azimuthal angular momentum $|m|=L$ by which numerical convergence can be reached. Being specific, we calculate for $N=2,3,4$ with $x=100$, $\widetilde U=4/\mu$ and $L=40$. 
%-----------------------figure--------------------------------------------------
\begin{figure}[t]
\begin{center}
\includegraphics[width=1 \textwidth]{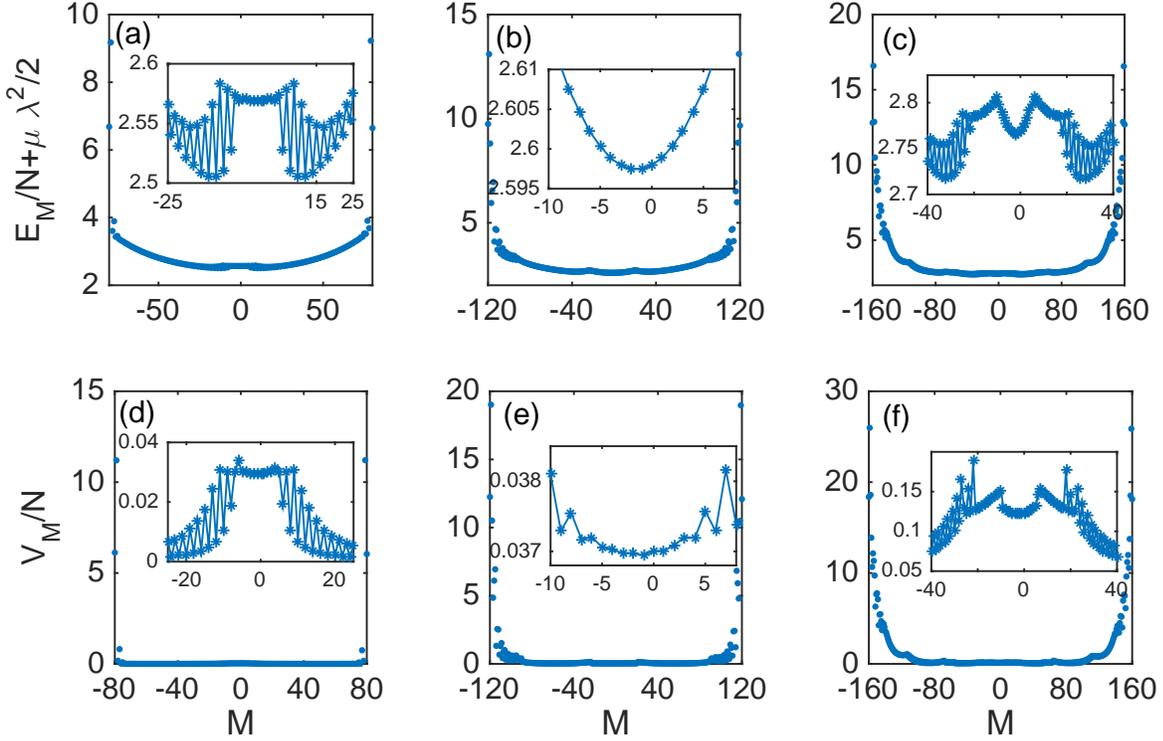}
\end{center}
\caption{(Color online)  (a)-(c) Lowest Energies $E_M$ (in units of $1/2\mu R^2$) of different total azimuthal
angular momentum $M$ for different $N=2,3,4$,
respectively. The inset is the enlarged view around the ground state. (d)-(f) $V_M/N$ (in units of $1/2\mu R^2$) for each state of $M$ for $N=2,3,4$ from (d) to (f), respectively. Here, we take $x=100$, $\widetilde U=4/\mu$, and $L=40$.}
\label{Figenergy}
\end{figure}
%-----------------------figure--------------------------------------------------

%-----------------------figure--------------------------------------------------
\begin{figure}[t]
\begin{center}
\includegraphics[width=1 \textwidth]{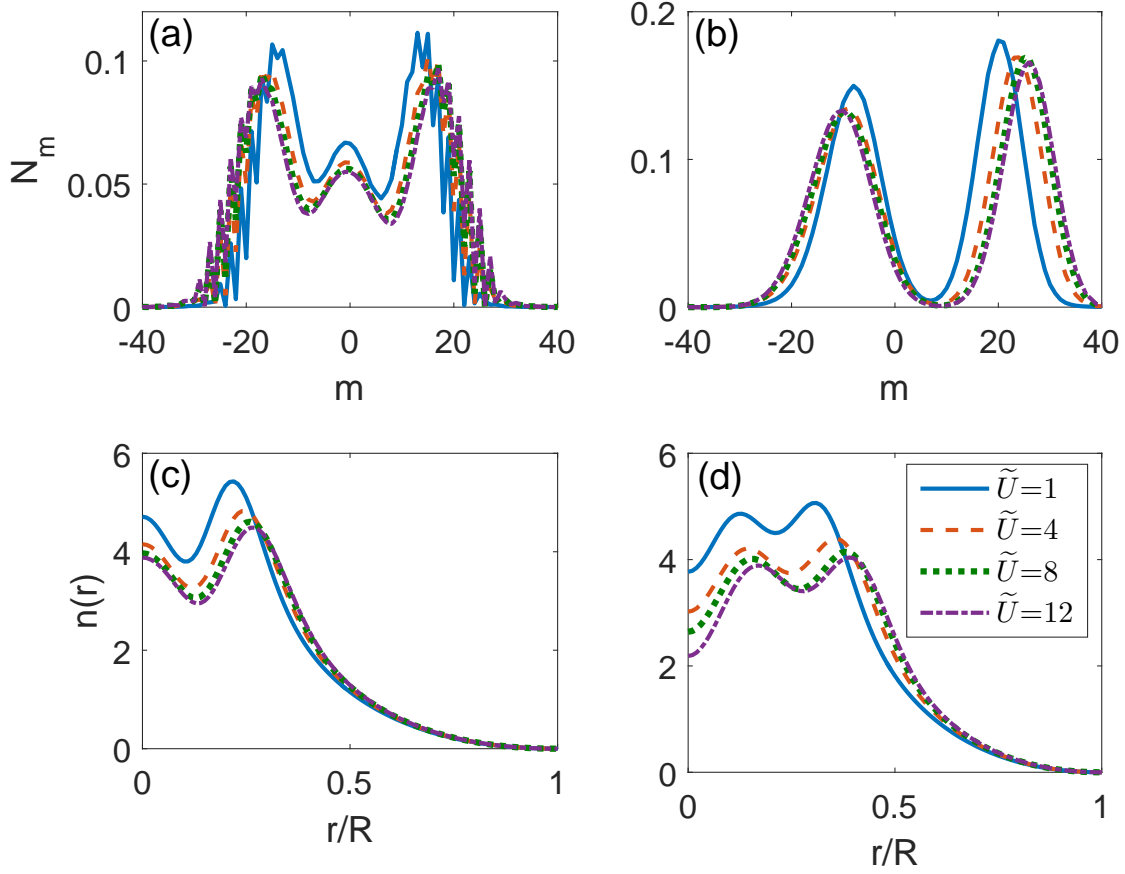}
\end{center}
\caption{(Color online) (a), (b) The ground state occupation number $N_m$ of the lowest band single particle states for different $\tilde{U}$.
(c), (d) Density distribution $n(r)$ of the ground state in real space for different $\widetilde{U}$. The left and right columns are corresponding to $N=3$ and $4$, respectively.}
\label{nm}
\end{figure}
%-----------------------figure--------------------------------------------------

Figure~\ref{Figenergy} shows the total energy $E_{M}$ and the expectation value of
the interaction Hamiltonian  $V_M\equiv \lr{H_{\mathrm{int}}}$ in the eigenstate of a cluster of bosons with total azimuthal angular momentum is $M$. There are
two degenerate ground states at $M=10$ and $-12$ for $N=2$ (Figure \ref{Figenergy}(a)), at $M=-1$
and $-2$ for $N=3$ (Figure \ref{Figenergy}(b)), and at $M=28$ and $-32$ for $N=4$ (Figure \ref{Figenergy}(c)).
The double degeneracy of the ground states can be understood in the following way. Without interaction,
if $\Phi_{m}$ of $L_z+S_z=m+1/2$ is a single particle eigenfunction with energy $E_m$ of Eq.~(\ref{eq1}), so is $e^{i\frac{\pi}{2}\sigma_y}\Phi_{m}^{*}(\mathbf{r})$. By using $J_{m}(x)=(-1)^{m}J_{-m}(x)$, we find that $e^{i\frac{\pi}{2}\sigma_y}\Phi_{m}^{*}(\mathbf{r})$ has $L_z+S_z=-m-3/2$;
both the single particle states of angular momentum symmetric about $L_z+S_z=-1/2$ are degenerate. The interaction Hamiltonian is time-reversal invariant with the time-reversal operator defined as $\mathcal{T}=e^{-i\frac{\pi}{2}\sigma_y}\mathcal{K}$ with the complex conjugation operator $\mathcal{K}$. Thus in the case of a cluster of $N$ bosons, the energy spectrum should be symmetric with respect to $M=-N/2$, which agrees with Figure~\ref{Figenergy}.

Figure~\ref{Figenergy} also shows that there is a large range of $M$ for which the low lying $E_M$ is rather flat. In the flat region, the energy per particle for the relevant $\widetilde{U}$ we consider is about from $0.1\times 1/2\mu R^2$ to $0.3\times 1/2\mu R^2$, measured with respect to the band bottom of the lower Rashba band. This is much smaller than the energy difference between the lowest two bands $n=0$ and $1$ at $m=0$, approximately $10/2\mu R^2$, and justifies our lowest band approximation. Whenever the energy per particle (with respect the lowest band bottom whose value is $2.47\times 1/2\mu R^2-\mu\lambda^2/2$) becomes comparable to $10/2\mu R^2$, our lowest band approximation breaks down.

On the other hand, Figure~\ref{Figenergy} shows that the interaction energy per particle of the ground state $|s_0\rangle$ is orders of magnitude smaller than its energy per particle. This magnitude difference suggests that the bosons are taking advantage of the nearly flat lowest band to correlate in a way such that the interaction energy, usually of order $\ln(\mu\lambda R)/2\mu R^2$ (see Eq.~(\ref{hprime}) below), which for $x=100$ is about $4.3\times 1/2\mu R^2$, is greatly suppressed. Figure~\ref{nm} (a-b) show that when $\widetilde U$ increases, bosons are prompted to the single particle states of larger $|m|$ in the lowest band, thus increasing the kinetic energy. Correspondingly, Fig.~\ref{nm} (c-d) shows that the density distribution is pushed away from the center due to occupation of higher angular momentum states $m$. Here $N_m$ is the occupancy of single particle state with angular momentum $m$. With the spreading of $N_m$, it is possible for bosons to explore a larger set of nearly-degenerate lowest band single particle states to suppress interactions. This difference also indicates that the kinetic energy of the cluster, which arises from the weakly dispersing band (whose width is about $0.6\times 1/(2\mu R^2)$ for $L=40$ and $x=100$), comprises the majority of the energy of the system. 
For example, the dispersion of the single particle lowest band for $x=100$ (cf.~Fig.~\ref{Figspdisk}(d)) can be well fitted by the formula $(E_m+\mu\lambda^2/2)/(1/2\mu R^2)=5.11\times10^{-4}(m+1/2)^2+2.47$. Given that the boson occupation number $N_m$ is noticeable up to $m=20$ (see Fig.~\ref{nm}(a) and (b)), the kinetic energy of each boson is estimated to be about $0.2\times1/2\mu R^2$ when measured from the band bottom, agreeing with our numerical calculations. Thus it is necessary to start with an effective Hamiltonian for which the transverse excitations away from the degenerate ring are included.

%-----------------------figure--------------------------------------------------
\begin{figure}[t]
\begin{center}
\includegraphics[width=1 \textwidth]{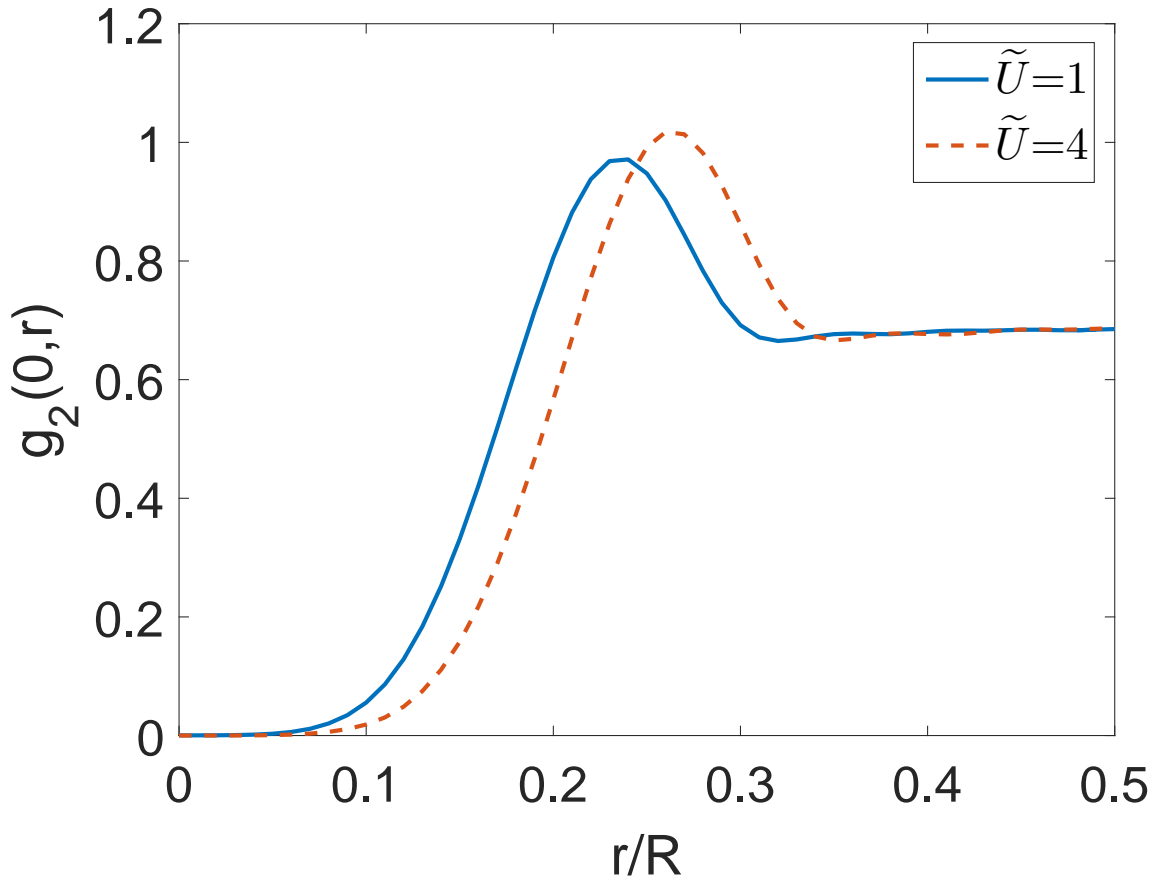}
\end{center}
\caption{(Color online)  Correlation function $g_2(0,r)$ of the ground state versus $r/R$ for different $\widetilde{U}$. The probability for two bosons to stay close to each other is suppressed and the suppression is larger as the increase of the interaction strength. Here, we show the case of $N=3$. The asymptotic value is about $2/3\sim 0.67$, differing from $1$ because of the finite number ($N=3$) of particles considered.}
\label{g2}
\end{figure}
%-----------------------figure--------------------------------------------------

%-----------------------figure--------------------------------------------------
\begin{figure}[t]
\begin{center}
\includegraphics[width=1 \textwidth]{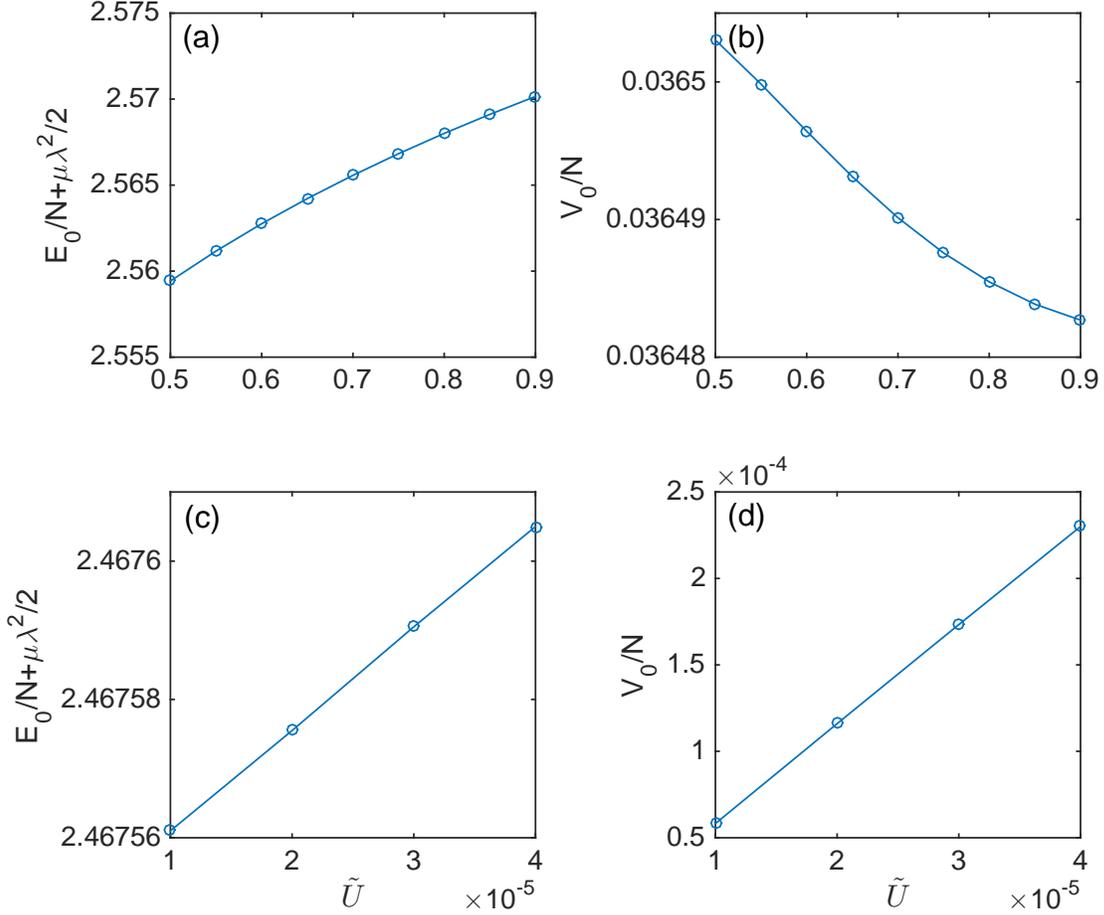}
\end{center}
\caption{(Color online) (a), (b) Energy $E_0$ and the interaction energy $V_0=\langle s_0|H_{\mathrm{int}}|s_0\rangle$ of the ground state $|s_0\rangle$ as functions of the interaction strength $\widetilde U$ in the strongly correlated regime, respectively ; (c), (d) $E_0$ and $V_0$ versus $\tilde{U}$ in the weakly correlated regime, respectively.
Here, we take $N=3$, $L=40$, $x=100$ and use $1/2\mu R^2$ as energy units and $1/\mu$ as units for $\widetilde U$.}
\label{Fig4}
\end{figure}
%-----------------------figure--------------------------------------------------

The correlation of the ground state $|s_0\rangle$ is manifested in the density-density function
\begin{align}
g_2(0,r)=\frac{\left\langle{:n(\mathbf r)n(0):}\right\rangle}{\left\langle n(\mathbf r)\right\rangle\left\langle n(0)\right\rangle},
\end{align}
where $n(\mathbf r)$ is the density operator of the bosons, the symbol $:\,:$ means normal ordering and $\langle\dots\rangle$ is averaging with respect to $|s_0\rangle$. Since $|s_0\rangle$ is an eigenstate of total $L_z+S_z$, $g_2(0,r)$ is a function of $r$. Figure \ref{g2} indicates that the probability for two bosons to stay close to each other is greatly suppressed at large interaction $\widetilde{U}$, and as $\widetilde{U}$ increases, such suppression becomes stronger. In Fig.~\ref{g2}, $g_2(0,r)$ shows a maximum at about $r/R=0.25$, which can be understood from the following semi-classical picture. From Fig.~\ref{nm} (a) for $N=3$, the most probable relative angular momentum between two bosons is about $\Delta m=40$. Due to the strong spin-orbit coupling, the typical linear momentum of a boson is about $\mu\lambda$ and the typical relative linear momentum between two bosons is about $2\mu\lambda$. Thus the most probable distance $\Delta r$ between two bosons is about $\Delta r/R=\Delta m/2\mu\lambda R=\Delta m/\sqrt2 x\approx0.28$. Beyond the maximum, $g_2(0,r)$ converges to a constant. The constant is noticeably smaller than unity due to the boson number $N$ being comparable to unity. Note that the numerator of $g_2(0,r)$ normalises as $\int d^2\mathbf r \left\langle{:n(\mathbf r)n(0):}\right\rangle=(N-1)\langle n(0)\rangle$, which we have checked numerically, while the denominator of $g_2(0,r)$ normalises as $\int d^2\mathbf r \left\langle n(\mathbf r)\right\rangle\left\langle n(0)\right\rangle=N\langle n(0)\rangle$. This means that for $N=3$, the asymptotic value of $g_2(0,r)$ is about $2/3\sim 0.67$, consistent with numerical findings.

The underlying correlation between the bosons also gives rise to an intriguing behavior of the interaction energy $V_0$ of the ground state $|s_0\rangle$ with respect to the interaction strength $\widetilde U$. Figure \ref{Fig4} shows the variation of the ground state energy $E_0$ and the interaction energy $V_0$ of the ground state $|s_0\rangle$ for $N=3$ with $\widetilde U$ both in the strongly and weakly correlated regimes. By the Hellmann-Feynman theorem, $E_0$ must be a strictly increasing function of $\widetilde{U}$, agreeing with Figure \ref{Fig4} (a) and (c). What is surprising is that, however, Figure \ref{Fig4} (b) shows that the interaction energy $V_0$ is a decreasing function of $\widetilde{U}$ in the range $\mu\widetilde{U}\in[0.5,0.9]$, contrary to what one would expect for an ideal condensed state (cf.~Tab.~\ref{Tab1}), including mean field vortex states. It is also opposite to the $\widetilde{U}$-dependence of the constructed wave-function $|\rm{Corr}\rangle$ (Eq.~(\ref{eq19})) and the super-fragmented state proposed in Ref.~\cite{QiZhou}. In the weakly correlated limit where the interaction energy is much smaller than the energy difference between neighbouring states which is found to be of order $10^{-3}$ for $x=100$, Figure \ref{Fig4} (d) shows that interaction energy $V_0$ increases linearly with $\widetilde{U}$. So does $E_0$ in Fig.~\ref{Fig4} (c). Thus as $\widetilde{U}$ increases, the ground state shall evolve from a weakly interacting regime where the usual mean field treatment is applicable to a strongly correlated regime where crucial interaction effects can not be captured by mean field.

\subsection{Effective low-energy Hamiltonian}
\label{cseffHam}

To understand why in the presence of strong spin-orbit coupling it is favorable for a cluster of bosons to correlate in a way to suppress the interaction energy, we extract the leading order of the interaction Hamiltonian Eq.~(\ref{eq27}). In the limit $x\to\infty$, we can simplify the matrix elements $V(m_1,m_2,m_3,m_4)$, and as shown in Appendix \ref{app2}, the final form of the interaction Hamiltonian takes a particularly simple form, to leading order in $\ln(\mu\l R)$,
\begin{align}
H'_{\mathrm{int}}=\frac{4\widetilde U}{\pi R^2}\ln(\mu\lambda R) \sum_{l}(A_{l}^{\dagger}A_{l}+B_{l}^{\dagger}B_{l}),\label{hprime}
\end{align}
where $A_{l}=\sum_{m}\widetilde{N}_{l-m}\widetilde{N}_{m}a_{m}a_{l-m}$ and $B_{l}=\sum_{m}(-1)^m\widetilde{N}_{2l-m}\widetilde{N}_m a_{m}a_{2l-m}$. The dimensionless normalization factor $\widetilde N_m$ is given in Appendix \ref{app2}. 
It is worth to emphasise that Eq.~(\ref{hprime}) is derived within the lowest band approximation, which as we argue above holds for our numerical calculation with $x=100$, cutoff $L=40$ and up to $N=4$. Obviously, for any states $\ket{\psi}$, $\bra{\psi}H'_{\mathrm{int}}\ket{\psi}\geq0$. The strength of $H'_\text{int}$ is logarithmically divergent in the large spin-orbit coupling limit. If we now neglect the weak dispersion of the lowest band which is of order of $1/(\lambda^2R^4)$ as compared with $\ln(\mu\lambda R)/R^2$, the ground state will be the one which minimises $H'_{\mathrm{int}}$. For a state $\ket{s}$, $\bra{s}H'_{\mathrm{int}}\ket{s}=0$ only if $A_l \ket{s}=B_l \ket{s}=0$ for any $l$. The existence of such nontrivial correlated states can be considered in the following way. We assume that the spin-orbit coupling is so large that the lowest band can be considered flat for the states of $m\in[-L,L]$. For $N$ bosons, the dimension of the Hilbert space is $(2L+N)!/(2L)!N!$. The requirement that $A_l \ket{s}=B_l \ket{s}=0$ for $l\in[2L,-2L]$ imposes $2(4L+1)(2L+N-2)!/(2L)!(N-2)!$ constraints. For fixed $N$, the constraints can be simultaneously satisfied if $L$ is large enough; in other words, when the spin-orbit coupling is large enough. On the other hand, an ideal condensate of the form Eq.~(\ref{eq17}) can not avoid the interaction of the form $H'_\text{int}$ since a condensate wave-function is determined by assigning $2L+1$ superposition coefficients. Thus in this limit, the ground state $|s_0\rangle$ of a cluster of bosons is expected to have the correlation such that
$\langle s_0|\sum_l(A_l^\dagger A_l+B_l^\dagger B_l) |s_0\rangle\ll1$, which is compatible as shown in Fig.~\ref{Fig4} (b) where the interaction energy $V_0$ is much smaller than $1/2\mu R^2$. It is worth mentioning that $H'_\text{int}$ is formally similar to a Hamiltonian whose ground state can be correlated Mott insulators, corresponding to fractional quantum hall states \cite{Lee2004,Lee2005}. The nonlocal nature of $\widetilde{N}_{m}$ (see Appendix \ref{app2}), however, precludes constructing exact ground states of $H'_\text{int}$ in our case. It is important to note the difference between Eq.~(\ref{hprime}) and other effective interactions derived in Refs.~\cite{Ozawa} and \cite{QiZhou} for infinite systems; the renormalised effective interaction derived in Ref.~\cite{Ozawa} is applicable to the lower branch states while the one in Ref.~\cite{QiZhou} is meant for states on the degenerate ring. In all, the above considerations based on Eq.~(\ref{hprime}) made it clear why other than an ideal condensed state, the bosons would prefer to be in a ground state with strong correlations, whose properties we have revealed in details by the previous numerical calculation.

\section{Conclusion}
\label{conclusion}

In this work, we describe how the enhanced low-energy density of states changes the properties of a spin-half boson subjected to the Rashba spin-orbit coupling. We carry out an exact diagonalisation calculation for a cluster of bosons with strong spin-orbit coupling and reveal the correlated nature of its ground state. 
We derive the corresponding effective Hamiltonian [Eq.(\ref{hprime})], which is particular simple in form and suggests a correlated ground state that is analogous to correlated Mott insulator and quantum Hall states, suggested earlier in the literature.

However, standing alone, the argument based on the effective Hamiltonian (\ref{hprime}) is deficient at least in two aspects. Firstly, the weak dispersion of the lowest band and sub-leading terms of the interaction Hamiltonian need to be considered in order to uniquely determine the true ground state. Secondly, the comparison with ideal condensed states is less relevant if bosons can lower their energy substantially by strong depleting from condensates. The later problem requires one to investigate possible variational ground states of a condensed boson system with strong depletion, which we do not attempt here; see however, the relevant discussion of Ref.~\cite{Radic2015}. 
%There are still a lot remained to be understood about the correlated ground state. Firstly, it is necessary to see the effects of subleading terms to Eq.~(\ref{hprime}). Secondly, 
More extensive numerical calculations ({\em e.g.}, variational Monte Carlo) are necessary to provide more evidences to the correlated nature of the ground state.

\section{Acknowledgment}
We thank W. Cole, N. Trivedi, A. Paramenkanti and J. Yao for their discussions in the initial stage of this work. 
Z.X. is supported by NSFC under Grant No. 11547111. Z.Y. is supported by NSFC under Grant No. 11474179 and 11204152, and the Tsinghua University Initiative Scientific Research Program. S.Z. is supported by Hong Kong Research Grants Council, GRF HKU-709313P, Collaborative research fund, HKUST3/CRF/13G, and the Croucher Foundation under the Croucher Innovation Award.

\appendix
\label{app}
\section{Single Particle States on a Disk.}
\label{app1}

To obtain the explicit form of the function $f_{nm}(r)$ and $\xi_{nm+1}(r)$, we note that without the hard wall boundary condition, the single particle wave-function can be written as, since $m=S_z+L_z$ is a good quantum number
\begin{equation}
\label{eq24}
\chi_{m}(r,\varphi)=\left(
\begin{array}{c}
\alpha e^{im\varphi} J_{m}(kr)\\
\beta e^{i(m+1)\varphi} J_{m+1}(kr)
\end{array}%
\right),
\end{equation}
with $\a$ and $\b$ is the coefficients to be determined later. Requiring that
\be
H_{\mathrm{kin}}\chi_{m}(k)=E_m\chi_{m}(k),
\ee
where $E_m$ is the eigenvalue with magnetic quantum number $m$. For a specific value of $E_m$, there are two values of $k$ that can be found, by requiring that
\begin{equation}
\label{eq25}
{\rm det}\left[
\begin{array}{cc}
\frac{\hbar^2 k^2}{2\mu}-E_{m} & \lambda k\\
\lambda k & \frac{\hbar^2 k^2}{2\mu}-E_{m}
\end{array}
\right]=0.
\end{equation}
As a result, we find that $\epsilon_{m}^{(\pm)}=E_{m}+\mu \lambda^2 \pm \sqrt{2\mu \lambda^2 E_{m}+\mu^2\lambda^4}$ with $\epsilon_{m}^{(\pm)}=\hbar^2k_{m}^{(\pm) 2}/{2\mu}$. Let us denote the corresponding wave-functions as $\chi^\pm_m(r,\varphi)$. The coefficients are given by
\begin{align}
\alpha_{m}^{(\pm)}&=\frac{\lambda k_{m}^{(\pm)}}{E_{m}-\epsilon_{m}^{(\pm)}}\beta_{m}^{(\pm)}.
\end{align}
At the moment, there is no requirement on the overall normalization. Note that $\alpha_{m}^{(\pm)}$ are functions of $E_m$. Now we need to impose the boundary conditions on the single particle wave-function such that it vanishes at $r=R$. This can be done by forming a superposition of $\chi^\pm_m(r,\varphi)$ (both of energy $E_m$), with coefficients $a_m,b_m$
\begin{eqnarray}
\label{eq26}
\Phi_{m}(r,\varphi)&=&a_m\left(
\begin{array}{c}
\alpha^{(+)}_{m} e^{im\varphi} J_{m}(k_{m}^{(+)}r)\\
\beta^{(+)} _{m}e^{i(m+1)\varphi} J_{m+1}(k_{m}^{(+)}r)
\end{array}%
\right)+b_m\left(
\begin{array}{c}
\alpha^{(-)}_{m} e^{im\varphi} J_{m}(k_{m}^{(-)}r)\\
\beta^{(-)}_{m} e^{i(m+1)\varphi} J_{m+1}(k_{m}^{(-)}r)
\end{array}%
\right)
\end{eqnarray}
Requiring $\Phi_{m}(R,\varphi)=0$ leads to the equation for the coefficients $a_m$ and $b_m$ and the condition for them to have non-zero solution is that the corresponding determinant is zero. Namely, we require
\begin{equation}
{\rm det}\left[
\begin{array}{cc}
\alpha^{(+)}_{m}  J_{m}(k_{m}^{(+)}R) & \alpha^{(-)}_{m}  J_{m}(k_{m}^{(-)}R) \\
\beta^{(+)} _{m} J_{m+1}(k_{m}^{(+)}R) & \beta^{(-)}_{m}  J_{m+1}(k_{m}^{(-)}R)
\end{array}
\right]=0.
\end{equation}
From this condition, we determine the allowed set of eigen-energy $E_{nm}$, which we label by the radial quantum number $n$. Once we know $E_{nm}$, we can find the corresponding $k^{(\pm)}_{nm}$ and coefficients $\a_{nm}^{(\pm)}$, $\b_{nm}^{(\pm)}$, $a_{nm}$ and $b_{nm}$ for a specific radial quantum number $n$. The explicit form of  $f_{nm}(r)$ and $\xi_{nm+1}(r)$ can then be written as
\begin{align}
f_{nm}(r) &=a_{nm}\a_{nm}^{(+)} J_m(k^{(+)}_{nm}r)+b_{nm}\a_{nm}^{(-)}  J_m(k^{(-)}_{nm}r),\\
\xi_{nm+1}(r) &=a_{nm}\b_{nm}^{(+)} J_{m+1}(k^{(+)}_{nm}r)+b_{nm}\b_{nm}^{(-)}  J_{m+1}(k^{(-)}_{nm}r).
\end{align}
The full spinor wave-function $\Phi_{nm}(r,\varphi)$ can now be chosen orthonormal with $\int d\varphi \int rd r \Phi_{nm}^{\dagger}(r,\varphi)\Phi_{n'm'}(r,\varphi)=\delta_{n,n'}\delta_{m,m'}$. For the numerical calculation shown below, we use $1/(2\mu R^2)$ as the units for energy and define $x=\sqrt{2\mu^2\lambda^2R^2}$ to quantify the strength of spin-orbit coupling.

\section{Analyzing effective Hamiltonian in limits}
\label{app2}
Consider insider a two dimensional cylinder of radius $R$. The single particle wave-function of
Eq.(\ref{eq1}) in the main text is
\begin{equation}
\label{Aeq7}
\left(
\begin{array}{c}
e^{im\varphi} J_{m}(kr)\\
-e^{i(m+1)\varphi} J_{m+1}(kr)
\end{array}%
\right).
\end{equation}
The wave-function are required to vanish at $r=R$; the single particle wave-function on the lowest band
in the limit $\lambda R \to \infty$ and under the open boundary condition is described by
 \begin{eqnarray}
\label{Aeq8}
\Phi_{m}(r,\varphi)&=&N_{m}\left[J_{m}(k^{-} R)\left(
\begin{array}{c}
J_{m}(k^{+}r)e^{im\varphi}\\
-J_{m+1}(k^{(+)}r)e^{i(m+1)\varphi}
\end{array}%
\right)
-J_m(k^{(+)}R)\left(
\begin{array}{c}
J_{m}(k^{(-)}r)e^{im\varphi}\\
J_{m+1}(k^{(-)}r)e^{i(m+1)\varphi}
\end{array}%
\right)\right] \notag \\
&=&N_{m}\left(
\begin{array}{c}
f_{m}(r) e^{im\varphi}\\
\xi_{m+1}(r)e^{i(m+1)\varphi}
\end{array}%
\right),
\end{eqnarray}
where $N_m$ is normalization factor. To the lowest order $k^{\pm}\sim \lambda \pm \sqrt{\Delta}$ where
$\Delta=E-E_{\mathrm{min}}\sim 1/R^2$ and set the mass of the particle $\mu=1$.

To determine the normalization factor $N_m$ and likewise in the following evaluation for the
interaction matrix elements, we expand
\begin{equation}
\label{Aeq9}
f_m(r)=-[J_{m}(k^{+}R)J_{m}(k^{-}r)-J_{m}(k^{-}R)J_{m}(k^{+}r)],
\end{equation}
\begin{equation}
\label{Aeq10}
\xi_{m+1}(r)=J_{m}(k^{+}R)J_{m+1}(k^{-}r)-J_{m}(k^{-}R)J_{m+1}(k^{+}r),
\end{equation}
to the lowest order of $\Delta/\lambda^2$ and find
\begin{equation}
\label{Aeq11}
f_m(r)=-2\sqrt{\Delta}[J^{\prime}_{m}(\lambda R)J_{m}(\lambda r)R-J_{m}(\lambda R)J^{\prime}_{m}(\lambda r)r],
\end{equation}
\begin{equation}
\label{Aeq12}
\xi_m(r)=2\sqrt{\Delta}[J^{\prime}_{m}(\lambda R)J_{m+1}(\lambda r)R-J_{m}(\lambda R)J^{\prime}_{m+1}(\lambda r)r].
\end{equation}

Thus the normalization condition is
 \begin{eqnarray}
\label{Aeq13}
\frac{1}{N_{m}^{2}}&=&2\pi\int_{0}^{R}d r r[f_m^2(r)+\xi_{m+1}^{2}(r)] \notag \\
&\sim&\frac{32}{3\pi\lambda^2}R^2\Delta[2-(-1)^m\sin(2\lambda R)],
\end{eqnarray}
where in the second line we use the asymptotic approximation of Bessel function of the first kind
$J_m(z)\sim \sqrt{\frac{2}{\pi z}}\cos(z-\frac{m\pi}{2}-\frac{\pi}{4})$.

To calculate the leading order of the interaction, we need to take into account only
\begin{equation}
\label{Aeq14}
f_m(r)=-2\sqrt{\Delta}J^{\prime}_{m}(\lambda R)J_{m}(\lambda r)R,
\end{equation}
\begin{equation}
\label{Aeq15}
\xi_{m+1}(r)=2\sqrt{\Delta} J^{\prime}_{m}(\lambda R)J_{m+1}(\lambda r)R.
\end{equation}
Let's define $\widetilde{N}_m=2\sqrt{\Delta}N_mJ^{\prime}_{m}(\lambda R)\sqrt{R^3/\lambda}$ which is of order one when $x\to\infty$.
The interaction matrix element is
 \begin{eqnarray}
\label{Aeq16}
V_{m_1,m_2,m_3,m_4}&=&\frac{2\pi\lambda^2\widetilde U}{R^2}(\prod_{i=1}^{4}\widetilde{N}_{m_{i}})\delta_{m_1+m_2,m_3+m_4} \notag \\
& &\int_0^{R}
drr[J_{m_1}(\lambda r)J_{m_4}(\lambda r)+J_{m_1+1}(\lambda r)J_{m_4+1}(\lambda r)][J_{m_2}(\lambda r)J_{m_3}(\lambda r)+J_{m_2+1}(\lambda r)J_{m_3+1}(\lambda r)] \notag \\
&\sim&\frac{4\widetilde U}{\pi R^2}(\prod_{i=1}^{4}\widetilde{N}_{m_{i}})
[1+\cos(\frac{m_1+m_3-m_2-m_3}{2}\pi)]\ln(\lambda R),
\end{eqnarray}
where
\begin{equation}
\label{Aeq17}
\int_0^R dr r J_{m_1}J_{m_2} J_{m_3}J_{m_4}\sim\frac{1}{2(\pi\lambda)^2}[1+\cos(\frac{m_1+m_4-m_2-m_3}{2}\pi)+\cos(\frac{m_1+m_3-m_2-m_4}{2}\pi)]\ln(\lambda R).
\end{equation}

The symmetrized matrix element is
\begin{equation}
\label{Aeq18}
\widetilde{V}_{m_1,m_2,m_3,m_4}=\frac{1}{4}(V_{m_1,m_2,m_3,m_4}+V_{m_2,m_1,m_3,m_4}+V_{m_1,m_2,m_4,m_3}+V_{m_2,m_1,m_4,m_3}).
\end{equation}
Thus
\begin{equation}
\label{Aeq19}
\widetilde{V}_{l-k,k,k^{\prime},l-k^{\prime}}=\frac{\widetilde U}{\pi R^2}(\prod_{i=1}^{4}\widetilde{N}_{m_{i}})
[4+2\cos(k^{\prime}-k)\pi+2\cos(l-k-k^{\prime})\pi]\ln(\lambda R).
\end{equation}
At this stage, if $l$ is odd
\begin{equation}
\label{Aeq20}
\widetilde{V}_{l-k,k,k^{\prime},l-k^{\prime}}=\frac{4\widetilde U}{\pi R^2}(\prod_{i=1}^{4}\widetilde{N}_{m_{i}})\ln(\lambda R).
\end{equation}
If $l$ is even,
\begin{equation}
\label{Aeq21}
\widetilde{V}_{l-k,k,k^{\prime},l-k^{\prime}}=\frac{4\widetilde U}{\pi R^2}(\prod_{i=1}^{4}\widetilde{N}_{m_{i}})
[1+\cos(k^{\prime}-k)\pi]\ln(\lambda R).
\end{equation}
Thus the leading order of the interaction Hamiltonian is
 \begin{eqnarray}
\label{Aeq22}
H'_{\mathrm{int}}&=&\frac{4\widetilde U}{\pi R^2}\ln(\lambda R)\left[
\sum_l\left(\sum_k \widetilde{N}_{l-k}\widetilde{N}_{k}a_{l-k}^{\dagger}a_{k}^{\dagger}\right)
\left(\sum_{k^{\prime}}\widetilde{N}_{l-k^{\prime}}\widetilde{N}_{k^{\prime}}a_{k^{\prime}}
a_{l-k^{\prime}}\right) \right. \notag \\
& &\left. +\sum_{l=even}\left(\sum_k \widetilde{N}_{l-k}\widetilde{N}_{k}e^{ik\pi}a_{l-k}^{\dagger}a_{k}^{\dagger}\right)
\left(\sum_{k^{\prime}}\widetilde{N}_{l-k^{\prime}}\widetilde{N}_{k^{\prime}}e^{-ik^{\prime}\pi}a_{k^{\prime}}a_{l-k^{\prime}}\right)\right].
\end{eqnarray}
The nice form of $H'_{\mathrm{int}}$ suggests we define
\begin{equation}
\label{Aeq23}
A_{l}=\sum_{k^{\prime}}\widetilde{N}_{l-k^{\prime}}\widetilde{N}_{k^{\prime}}a_{k^{\prime}}a_{l-k^{\prime}},
\end{equation}
\begin{equation}
\label{Aeq24}
B_{l}=\sum_{k^{\prime}}(-1)^{k^{\prime}}\widetilde{N}_{2l-k^{\prime}}\widetilde{N}_{k^{\prime}}a_{k^{\prime}}a_{2l-k^{\prime}},
\end{equation}
and have $H'_{\mathrm{int}}=\frac{4\widetilde U}{\pi R^2}\ln(\lambda R)\sum_l (A^\dagger_lA_l+B^\dagger_lB_l)$.

%-------------------------

\end{document}